# Performance report of heuristic algorithm that cracked the largest Gset Ising problems (G81 cut=14060)

**Kenneth M. Zick**[1]
[1]University of Southern California – Information Sciences Institute, Arlington, VA 22203 USA
Corresponding author: Kenneth M. Zick (e-mail: kzick@isi.edu).

**ABSTRACT** For the past 25 years, the Gset benchmark problems have challenged all manner of Ising and Max-Cut solvers. The largest of these problems have remained unsolved by any heuristic algorithm. In this report we provide data showing dramatically better speed and accuracy on these large sparse problems. Our newly discovered heuristic algorithm called Cosm reaches high (99.9% of best) solution quality orders of magnitude faster than the previous best heuristic solver results. Additionally, when afforded enough steps Cosm attains higher cuts than ever previously reported, specifically on instances G72 (cut=7008), G77 (cut=9940), and the 20,000-variable G81 (cut=14060). This report includes solution bitstrings so that the cuts can be independently validated. Remarkably, the new best solutions appear to be optimal. We believe the results are an early hint of disruptive opportunities for unconventional, hardware-centric approaches to algorithm discovery.

## I. INTRODUCTION

An inordinate fraction of the world's computations have the same general character: optimization of a function. Whether real-time decision making in UAV swarms, design of billion-transistor electronic circuits, or training of AI models in enormous data centers, there are widespread needs in defense, science, and commercial industry for solvers that can quickly attain high-quality solutions to high-dimensional optimization problems. These problems are often categorized as either discrete or continuous, though innovative methods targeting one category can sometimes be imported into another (semidefinite programming, gradient descent, etc.). In binary quadratic optimization, the Gset problems [1] were created between 1999 [2] and 2000 [3] and have been used ever since as challenge problems for a wide range of Ising machines and solvers of Ising/Max-Cut/QUBO/UBQP problems. The three largest Gset problems {G72, G77, G81} have 10,000 to 20,000 variables, a toroidal square grid structure, and a weighted Max-Cut formulation with random weights $w_{ij} \in \{-1, +1\}$. Given the toroidal structure, such problems are NP-hard [4]. To convey the scale, the structure of the largest problem is illustrated in Appendix A. In this report we provide data showing that dramatically better heuristic performance on such problems is possible, and we provide three bitstrings of best reported solutions for independent validation.

## II. BACKGROUND AND PRIOR WORK

The Max-Cut objective function can be expressed as

$$\max \frac{1}{2} \sum_{1 \le i < j \le n} w_{ij}\left(1 - x_i x_j\right) \qquad (1)$$

with $n$ binary variables $x_i \in \{-1, +1\}$ [5]. Alternatively, the large Gset problems can be thought of as Ising problems with $J_{ij} = -w_{ij}$, spin variables $s_i \in \{-1, +1\}$ and the Ising Hamiltonian to be minimized

$$H(\boldsymbol{s}) = -\sum_{i<j} J_{ij} s_i s_j - \sum_i h_i s_i \qquad (2)$$

where each local field $h_i$ is 0 for these instances.

One heuristic algorithm sometimes cited as a point of comparison [6] for the Gset problems is Breakout Local Search [7], though its speed and accuracy on the largest problems has since been surpassed. Toshiba's landmark Simulated Bifurcation Machine heuristic (SBM) has demonstrated state-of-the-art performance on certain Ising problems [8]. On the largest Gset problems, SBM topped out at ~99.6% solution quality with a time-to-target of more than 10 hours [9][10]. A recent study evaluated quantum annealing-inspired heuristics and four variants of a simulated Coherent Ising Machine, and found that the G81 solution qualities were all well below the best [6]. In terms of exact solvers, there have been some recent advancements in handling sparse problems [11][12]. The commercial solver Gurobi has recently been used to solve G72/77/81 [13] (as discussed in Section III).

Only a single heuristic algorithm of which we are aware has reported high (99.9%) levels of solution quality on these problems: Teams of Global Equilibrium Search-Path Relinking Algorithms (GES-PR) by Shylo, Glover and Sergienko [14]. On G77, one of the solver teams reached 99.9% solution quality on eight of 20 trials. On G81, a team reached 99.9% on 1/20 trials. The implementation used an Intel Core i7-3770 CPU @3.4GHz and 8GB of RAM. Given the measured run times, the times-to-target to reach 99.9% solution quality (with 99% probability) on G77 and G81 correspond to 7 hours and 77 hours, respectively.

## III. EXPERIMENTAL RESULTS

**Methodology.** We implemented an iterative heuristic algorithm called Cosm in a MATLAB-based proof-of-concept. (The algorithm is to be published shortly after this report.) We tuned the algorithm parameters via testing on a 5000-variable problem (Gset G57), then conducted solver campaigns for each of the three largest Gset problems {G72, G77, G81}. Solutions were read out at the end of each trial. The proof-of-concept (referred to as Cosm-MATLAB) used MATLAB R2024b and a laptop computer with an Intel Core Ultra 7 155H CPU @3.8GHz, 16GB RAM, and 6 MATLAB workers. This implementation is a successor to the earlier implementation for which results were posted in Ref. [10]. In the solver campaigns, the only parameter that was tuned for each instance was the number of sweeps per trial (aka trial length). Cosm was run for 100 trials for each instance. Given the success probability $P_S$ of reaching (or exceeding) a target solution quality, the number of repetitions $r$ required to reach the target with at least 99% probability is taken to be

$$r = \max\left(1.0, \frac{\log(1-.99)}{\log(1-P_S)}\right). \tag{3}$$

We refer to the fundamental step in Cosm as a sweep, and define sweeps-to-target (STT) and time-to-target (TTT) as

$$\mathrm{STT}(target) = S_{trial} r \tag{4}$$

$$\mathrm{TTT}(target) = t_{trial} r \tag{5}$$

where $S_{trial}$ is the number of sweeps per trial and $t_{trial}$ is the average execution time per trial.

**Highest solution quality attained.** For all three problems {G72, G77, G81}, Cosm found higher cuts than reported by any other method. First, for G72 and G77, Cosm reached cuts of 7008 and 9940. The success probabilities with the current Cosm-MATLAB campaigns are significantly higher than in our previous report [10], mainly due to an improved setting of a single solver parameter. As a result, the sweeps-to-target is 5–6x better. The problem instance used to tune the parameters (G57) is likely more representative of the largest Gset problems than what was used for tuning in the previous Cosm campaigns. Surprisingly, for the largest Gset problem G81, Cosm found better solutions here (14058 and 14060) than all previously reported values. The highest and average G81 cuts achieved by Cosm are depicted in Fig. 1 for a variety of trial lengths. The figure illustrates that Cosm achieves very high G81 solution quality even for short trial lengths, and with increasingly longer trials relentlessly finds better and better cuts all the way up to the best known.

Solution bitstrings for all three problems are provided in Appendices C, D, and E to allow the solution qualities to be independently validated. Validation is straightforward—plugging the variable values into objective function (1) with the G72/77/81 problem coefficients gives the best known cuts 7008, 9940, and 14060, respectively. The associated best reported Ising energies found by Cosm are −14022, −19672, and −28086.

**Optimality.** According to Ref. [13], the exact solver Gurobi has been able to certify the optimal solutions for the toroidal Gset problems. The solutions and time-to-solution have not been released as of this writing. It is our understanding and expectation that the best Cosm solutions will indeed match the Gurobi-certified optimal solutions once published, though this awaits official confirmation. If this is confirmed, it will represent a further significant milestone for Cosm. One other note regarding optimality. On instance G81, the second-best cut (14058) is at 99.986% solution quality which is near a common threshold for optimality [15]; Cosm reaches at least 14058 with high probability (32/100). The cuts across 100 trials are consistently high and range from 99.94% to 100%, as shown in Appendix B.

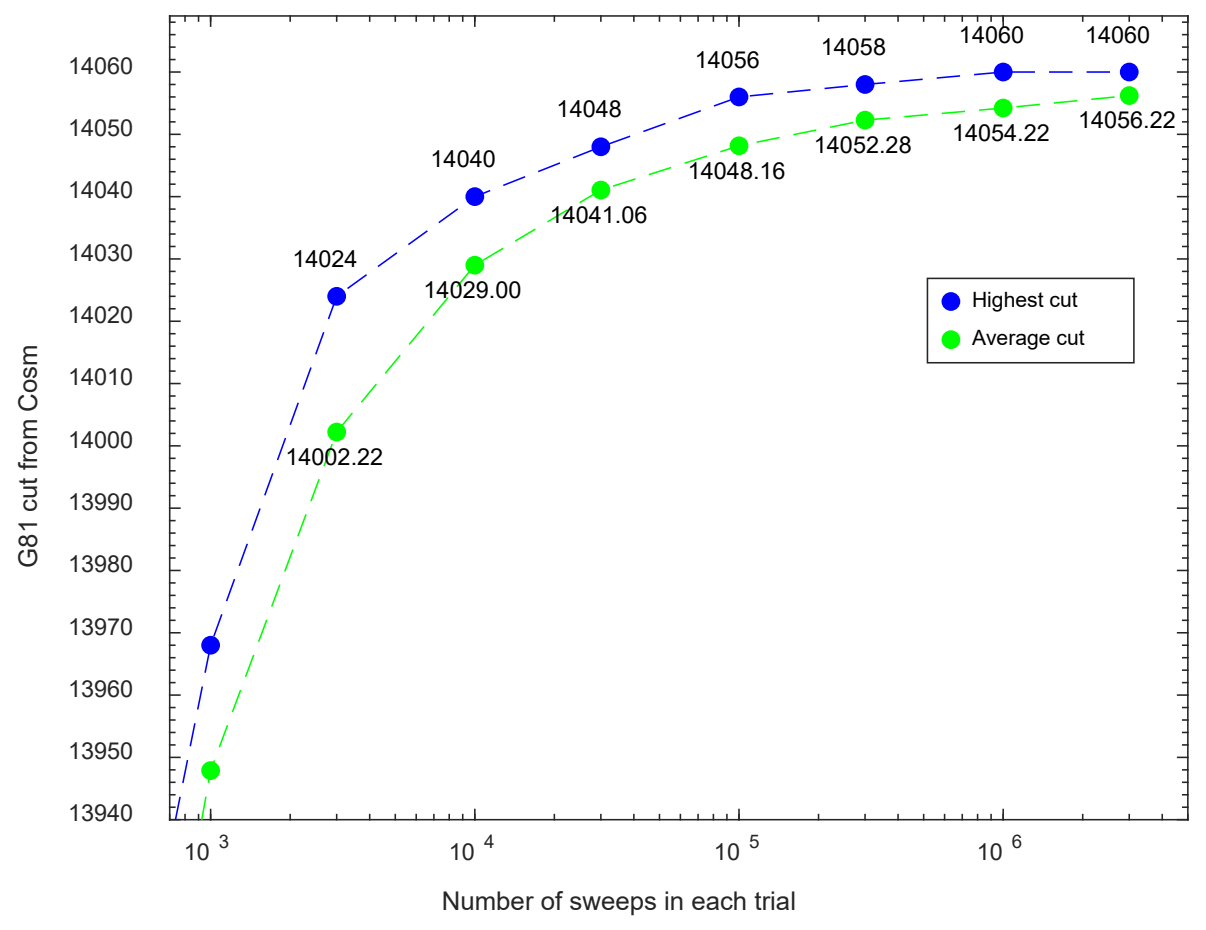

**FIGURE 1.** Highest and average G81 cuts found by Cosm, as a function of the length of each trial. Each pair of highest and average cut is from a 100-trial campaign. The numbers of sweeps are 1e3, 3e3, 1e4, 3e4, 1e5, 3e5, 1e6 and 3e6.

TABLE I
COSM PERFORMANCE DATA SHOWING SPEEDUPS TO 99.9% AND THREE NEW BEST REPORTED SOLUTIONS

| Gset problem instance | $\|V\|$ | $\|E\|$ | Target solution quality | State-of-the-art TTT from GES-PR [14] (s) | Number of Cosm sweeps per trial | Cosm success probability | Cosm sweeps to target | TTT of Cosm-MATLAB (s) | Speedup of Cosm-MATLAB TTT relative to GES-PR TTT | Projected TTT for parallelized Cosm implementation assuming 2ns/sweep |
|---|---|---|---|---|---|---|---|---|---|---|
| G72 | 10,000 | 20,000 | 99.9% (≥7000.99) | Not enough data | 80,000 | 55/100 | 577,000 | 47.5 | n/a | ~1.2 ms |
| | | | 100% (**7008**) | Did not reach | 1.5M | 34/100 | 16.6M | 1650 | n/a | ~33 ms |
| G77 | 14,000 | 28,000 | 99.9% (≥9930.06) | 25,800 | 80,000 | 66/100 | 342,000 | 39.4 | **655x** | ~0.7 ms |
| | | | 100% (**9940**) | Did not reach | 2M | 21/100 | 39.1M | 5320 | n/a | ~78 ms |
| G81 | 20,000 | 40,000 | 99.9% (≥14045.94) | 276,000 | 100,000 | 86/100 | 234,000 | 77.5 | **3560x** | ~0.5 ms |
| | | | 100% (**14060**) | Did not reach | 3M | 3/100 | 454M | 87,600 | n/a | ~910 ms |

**Speed.** The Cosm-MATLAB solver is a proof-of-concept simulator and not intended to be a production solver, though the simulator itself is quite fast in relation to the only other heuristic to achieve high solution quality. Whereas the state-of-the-art GES-PR solver needed 7 hours for G77 TTT to 99.9%, Cosm-MATLAB needs 39 s, a speedup of over 600x. For G81, GES-PR needed 77 hours for G81 to 99.9%, while Cosm-MATLAB needs 78 s, a speedup of more than 3000x. More interesting is the prospect of implementing Cosm not with MATLAB and a laptop CPU but with a highly parallelized implementation leveraging a GPU, FPGA, ASIC, wafer-scale engine [16], or other compute platform. While the hardware prospects are outside the scope of this report, we see a clear path to realizing an average sweep time of 2 ns. Given the sweeps-to-solution demonstrated in simulation, we project that a parallelized Cosm implementation would be able to reach 99.9% on any of the three largest Gset problems with a TTT of just ~1 ms. Such speed would be unprecedented. Likewise, with such an implementation the projected TTT to 100% of the best known solution quality would be just hundreds of ms.

Experimental results are summarized in Table I, including speedups in reaching 99.9%, the new best reported cuts, and projections of a parallelized Cosm solver.

## IV. CONCLUSION AND FUTURE WORK

The largest Gset problems have bedeviled Ising and Max-Cut solvers for 25 years. In this report we have provided data demonstrating much higher speed and accuracy, and have provided three bitstrings of best solutions for independent validation. Our collaboration plans to publish the Cosm algorithm shortly after this report. The algorithm can then be studied, enhanced, and perhaps applied to other problem types [17][18][19][20][21]. More importantly, Cosm as a case study points to broader, disruptive opportunities. Future work will seek to systematize our envisioned hardware-centric methods of discovery of breakthrough compute capabilities.

## ACKNOWLEDGMENT

The author thanks David Ferguson, Ryan Epstein, Bryan Jacobs, Itay Hen, Helmut Katzgraber, Nikhil Shukla, Matt French, Isidoros Doxas, Alex Marakov, Federico Spedalieri and Mohammad Sakib for helpful discussions; and Northrop Grumman Systems Corporation for support.

## APPENDIX A
## Gset G81 Problem Structure Visualization

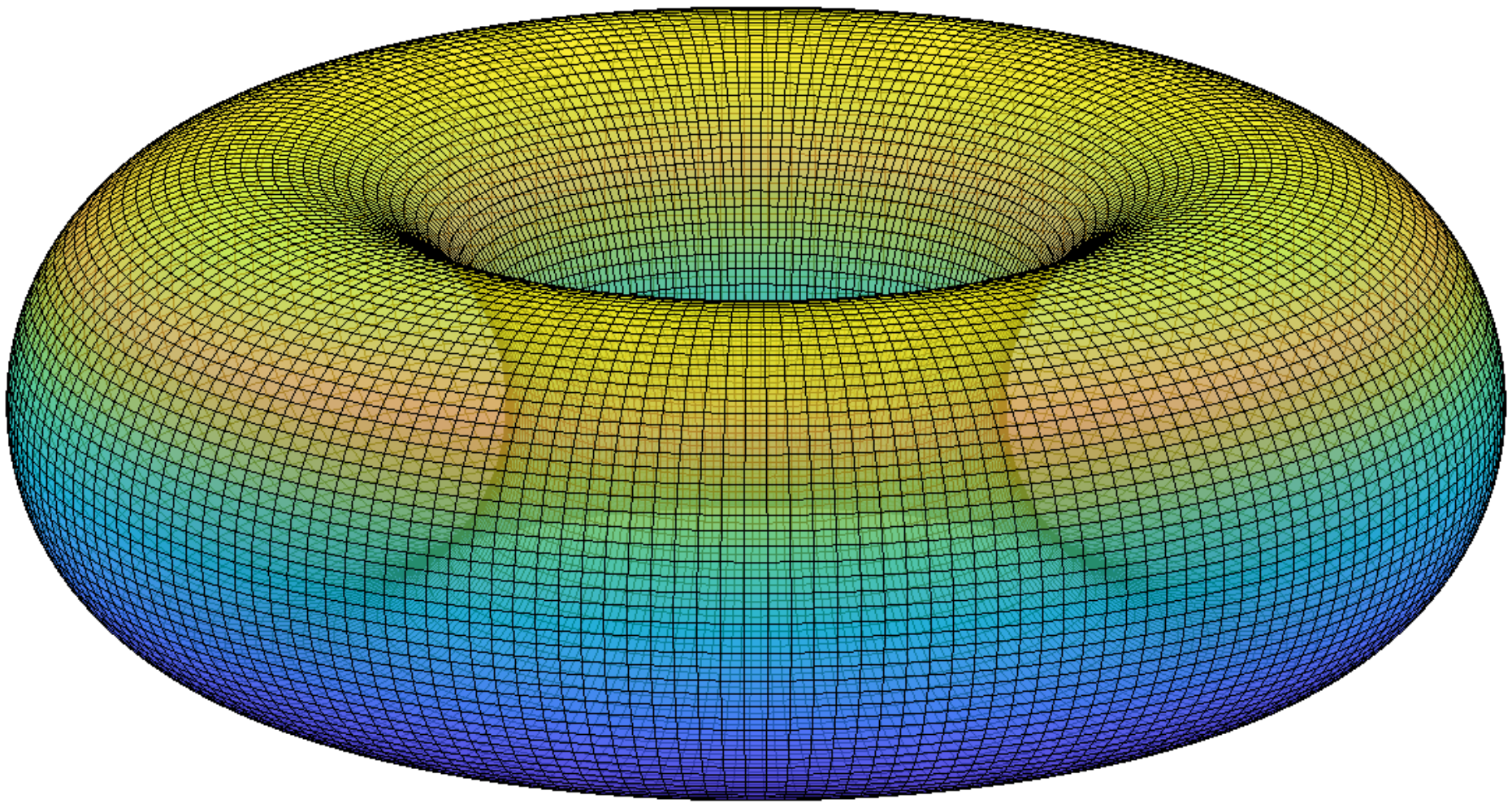

**FIGURE 2. Visualization of the G81 problem graph which has a toroidal square grid graph structure with 100×200=20,000 variables. Lines represent graph edges and intersection points represent variables. Each variable is coupled to four nearest neighbors. Due to the periodic boundaries in both dimensions, the problem is considered NP-hard [4].**

## APPENDIX B
## Cosm solution quality on problem G81

Some of the historic G81 milestones and other notable results are listed in Table II. The distribution of cuts for 100 trials (with 3M sweeps/trial) is shown in Fig. 3.

TABLE II
HIGHEST REPORTED AND OTHER NOTABLE G81 CUTS

| Solver | Highest G81 Cut |
|---|---|
| Cosm – 2025 (this work) | 14060 |
| GESPR – 2017 [22] | 14056 |
| GESPR – 2015 [14] | 14048 |
| PF-ESL – 2022 [23] | 14038 |
| MOH – 2017 [24] | 14036 |
| Breakout Local Search – 2013 [7] | 14030 |
| Toshiba SBM – 2021 [9] | 13992 |
| Rank-two relaxation - 2002 [5] | 13662 |
| SDP/dual scaling – 2000 [3] | 13448 |

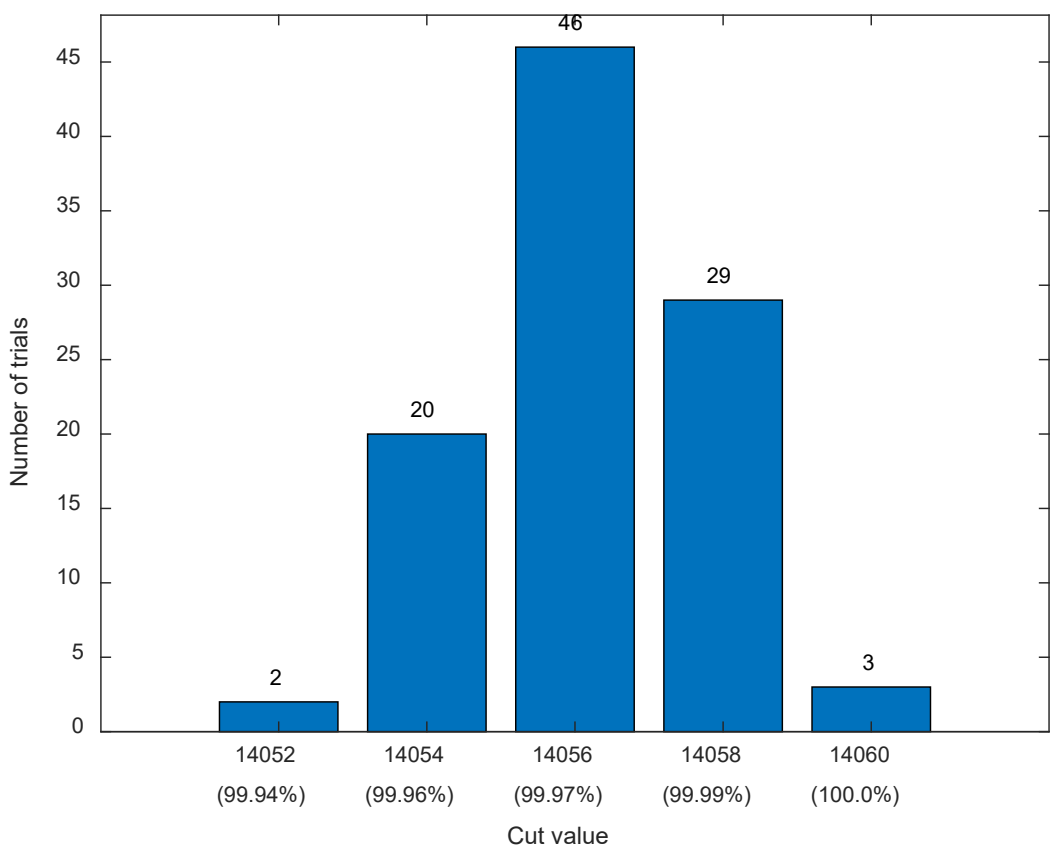

**FIGURE 3. Histogram of G81 cut values produced by the Cosm simulator. Solution quality vs. best reported is shown in parentheses. 100 trials with 3M sweeps/trial. The cuts of 14058 and 14060 are being reported here for the first time; the highest previously reported G81 cut value was 14056.**

## APPENDIX C
## G72 SOLUTION BITSTRING, CUT=7008

A Gset G72 solution bitstring found by Cosm is provided below. The cut value is 7008. The solution can be verified by expanding the hexadecimal string to binary values representing variables 1 through 10,000, mapping the $\{0,1\}$ values to $\{-1,+1\}$ (where 0 gets mapped to $-1$), then evaluating the weighted Max-Cut objective function (1) using the mapped values and the G72 problem instance defined by Ref. [1].

2b09d30989fc1e92a2b0e38060caaa99a0cbe4d24cfb995cb7b3d761a994892232981228c3de9c27b5e6160335d355166794a8b801e4ba42d5
9005799d5707261ca143895f6c282703ef380f73b4c116ddef92eb3db30e49fb83760f81462c17809ea4ed9cfe691087db098b7d63b6195c27c
22673d47638910cdeb70273f28e6f92d7bec9f3ac32ce83043772f48d3f0ff2d715fb64d02056e589be618e696d7fce1e96a133c7d23f0665e6d
5fae07ec8518a5578e4e45ae9537eb4dca9b0bf7f76d87966bf6fc37cb612ce479f8fe40ca1a13bdecc0950f83a65647c5f6bd8d3fd1f76df2e204
d89925d26cb94e14d57a387f18bfa957c6b150daf1c00e38193ddad2a0782aa8e77af3f3bf60c1ab1cc67d9688e9a4a374f0fa04d51699eb2d85
5a4a06064b30b827cdf193de7a50762b2020ad65341ec728587b07c86d471edc1378595e96cccae5d09bbfe9e8cc7a61a4e71b953086f45b5bb
536e5cbb25ab515319d4e87e9e6d8eb1028b63a6c7f84557aa103824e4f4408b662effe8c6c4d0d587e73beee61cbbe2444b6bf6d5d0ade83d6
1c9aa44b73f99eb742096d28e6487b53a137a5d7d1e7e415cc9e6175951b2b79cc137a46e23d1fde2e25aacc3ad7463da1ec6cf2943f4127b32
7d9e0d02cbe781e5ecdee24ae3080e759224f6f09a55f28cc64e682ee8e787aebd19d293881ce79d2e2c162db990ee049c0d85829a0f095cd777
e94e589d69bc96f44e7aa9f9f8ed8d707b809111c810cdd439cc47c9f7ff2257d3b0c33a9ea8d62e0f009462fca46b5dcf8026ee97d710062483f
ad50cd5fee90a6e48c76fc745b57d28e461c4d8a300a5e6aa833d4967c7fc8bd19a90891334da0cec200607b073e32e3715d5c77e0bfbfb44c86
2e58d1433c981e21c47aac4df80a6c935595c2d87bb1421c20e70bae2e4a261d73222e760ec8e9f5e90d27ceacc9f3af065385186481d99f6077
94f84ca409c0927ddb97994fa51febd7ba68a7e54caed7ccba2397e7d78ab96ee8089c4dfb10f6f5ccd96762c84f46d1d5727bc8bbd52d23ac5d
d8c03542d1ff1fa7fc1478fc77f4621ac73934925e0b63aa8001f21867f48baf4fb294717ef9ca6afeaaeac9f120f069575792c091a9fad1090a4b9
63fed4fb66b2b55eceb0e46f0ed515bc5722fb6977d091fcd5763e897b43e55b8702e8981638de22d9a900ae2f6184ef720455b3da9e8b4fc259
e92ab5493dd1e051d15015bdfee783115c6e5a4eb93526e641e2f0c66690d52310c1d98b3dc9ab09ffe1af9dac2b71a77725fef76375bcff53055
5f85144bb4aafc386363791a08fd828e233732e3bc1b78f9e9ac88f7fa2963bb025029c1c9809ab0ca5a366a5778907d03021fb5880427440a45
584bee0f4a1f51c4054534d303e6ee804d999b7952808e1266404e622e1b683bbce4ec8cdc25846e559acc1f27b8bbd579d8cae272c4eedaa9e
04eecdc3d12b901e7c4d485ff1cdd25889fa519ac0a85d967d949952942d617182878833310ae1709eaf5535dd0075a04063e0d2b64e6e1bac4
288b4ac58c63a0240facb756d9a221bce92a9f37f668a47b999c974542c9fa7a84ea857062e6fe5676c69ffd21b1ea83c1dc8f078769c91547119
8d456ad54dab797eef231198844ca0cb8cb690d266c4c20f9577e53140e0f6f4eeefd82b65b7e67c13a5beaf8287a2047db212f1943fae074d9a
b9e6a5df9d2a69db881e0a9cc20883a79a003edc5095afb5d92997c2fc460430

## APPENDIX D
## G77 SOLUTION BITSTRING, CUT=9940

A Gset G77 solution bitstring found by Cosm is provided below. The cut value is 9940. The solution can be verified by expanding the hexadecimal string to binary values representing variables 1 through 14,000, mapping the $\{0,1\}$ values to $\{-1,+1\}$ (where 0 gets mapped to $-1$), then evaluating the weighted Max-Cut objective function (1) using the mapped values and the G77 problem instance defined by Ref. [1].

c4ac172aa225affe46b7dc058dfc8f15b8da1f006c84cf8d1efd1e27b2e91ee8bfe955392ba99b12c4114371530451 1b4988bb4d6fa8527e32026
14f29525f0627a79bdae43efb060e9a2dda2fdc5087140d6b726f36d61f275b88b8dce388c1c12660262f42c1ef045a0f8f7012c2d59603444b3
856fdb29a7715082496aec6a70f52679b5c3480b7e254ccd18dab06e9334a5ff03a686205e0971870effac4e141376afd93fb9539edefefa82369
6c3426b3972ade3bba77d28430cc22937dfedecf212d43babf97c0fbda1704f09ecc07da67a6ee1290b11bc5bde12d7467a5418bb9090bebf64e
f34a1f05b278200df7cda881e1d8ad72bb986438c8ab5566bc18ab81d064459f1e7e6cb3738c1852c3012ab9d7cdecb22cb154b2bf33db0cc8d
e0dd78e1601d93fab354f4cb474024ec3969dd5f6f128fd379825abb100b121a986b72a4d1116e6748b795714c53ebf8e88d2e608eccc4d1843
288c290ed33d06348fc0327fa06c89e163fce5619b2c087bce0d1316440ad22f1db209cec93d7fb98a69487176488ce171110cc76cc5d5c33afd
ea201fa873cb0cfb6766a54ff835008c6ce3a2ea16c016630cabf506094833f465066719bd8dd489bf8625b8a686c6315e3aff089ffee7d00afde7
1f19728d5409495d046747f09ad826e945cfaee0e72ef9b46b0b290d0afea748646cb3718cac6ecb39bee38cd890ba80eb3c7356b083c2fb1583
271b6fce99e21dac865ab380cbbdda31e63069c3552af1d4696506c4dbe9ab8402d332cfa9fb730710135cdc283d44af7a33d260891ea86649a
6664f36e728b920dbc361917c8b6acd7229fc91f8860c27d6f7f872117667b6af2fe831ad40a519dd76b4e6cb244c301a1c44d97b5ebd6521a4c
400c9a2c180cdc2fb7e6e86eb6fd7092a9d41bcdd6b6b2ce7b5076b0987b93db05257e660a15c69a2e3679362ae2112b36c590cfffa74472731f
28ca1051093cc283a62557f2e24856a5f1cd3960e330e720a333734a26449bcfc6e7bfd3bb18e3c45a4dc65d7c82d58c5a79f11f995dc108ef65
76135e8efdaaf60f5af2f21fac39f8769e10308e20208ad040d80bc34154eca0c2419f8f2d5a5264ec347a11978ddb7fc62e961f57a42f2a8dfe4e
c591a4a8177e345ae6bde15bdcfcf812c87458b05599171c248c46d53dc175a5a24750aec235169ffcb7883df35ca3bd71643d002762cbeb61bf
7fb8f94fd7bf48fb5a3c0289cce930550e348b587dc65f6c9289021dbe72da98609f30292674005a4d09fb9053eeba8af493e11ab32e89cfbf312
8ff33b1ed4ae571db33d6f927134e3cbf8b6e21d87554e161396863873b95875fa3ff5b3f04ae06d3c9a6628e3fb952b04b5647fce7671250e4fe
5ce6e75e17ed0a7b1ddb902801d2558b75f734db70a2e5a288b361dbcfaaf9590e6939e8335e3f905b344b09f7e46469477e30f886cf25708e68
878c8b5de629b1b0cd8de749b9bf2f74c1332dfb2c30f79bf82568285d25ce629fabcc22a88f6206100d86e10e2fc322f3d446aea90afbb2068eb
bd88bb4038669987409bcf705f8b120d99763fa527d087188bfe7f4136b531c56f09eddab772d690ac2eedb0a25cf828c65ca8e835390c5f50bc
082cc2df571f6db0f86b65e7f932180ee9232c0416ec09e0bd7f349f9ded6f5a5a82c05168a4ccf6c21336c85e7cba6176ea9ba411a605ca007bb
178b263d50c6a61647aed28d2254b60a1dc409a1a9680174f2dc92720ddbff7e0d2a9a17de8b37940be2217119d96afffe17a84058ca6488925
b41a0cfe07c7f6c23bd9b585c406e103e4aeb28845df6ae7d3923a390c7146354498fd839b2029e8dab5580d723249d014ec8ec6314aba6e5bf
b003fea7952f8db32da3ab8c7da331ac43ee4830bfb71b7f9c7637bfaa00687bae24583cdb0fcf46695e18c3d2de0288e584703b763a7382252a
8fa472c9d7734c7bff14b646c15703b407459535c74d9bbc7c0e610e93f5bfdddb55751daad2ee42f8701f93abc3721e45b00b0515dfd288d556
70d69a010cd179724b36bd02e9607ed3315d364bb9bf26f60900bb739baa62d3ce398ca0c5b747427789a2ef829adabfefea33eb0a98e496bdc
6ffc4454c323211782aa65eabbf77ba8daa543b664bb261b99efa1144c537370c79f91bda95c3afb60570eb19edd15a968af79bf973da5ebbfb97
30d8bb33374e4b191a0d6d6d5f62d778a618f4d37690fa31e3171d4e97a6b5689d970b0189ab2347c2c62d9cd3c28502c761ef7429ab3d4269
8fb72c82a11695f5ce74157021fc16ea8bd9326d20af8262628a71fc68103f920c3da66e38f1f646f4ae7cffbaad760e4a0ccb20d4b59901b3cec6
21eab211e2739db1a9c2bbb87f26d5fb8531275955ad93b8bfed69e57b7b2f511f57579c370c82f60ccbc5999f67e486a4b0323d02aea73a8cf6
ed49b2e5d52e291

## APPENDIX E
## G81 SOLUTION BITSTRING, CUT=14060

A Gset G81 solution bitstring found by Cosm is provided below. The cut value is 14060 which is higher than previously reported values in the literature. The solution can be verified by expanding the hexadecimal string to binary values representing variables 1 through 20,000, mapping the $\{0,1\}$ values to $\{-1,+1\}$ (where 0 gets mapped to $-1$), then evaluating the weighted Max-Cut objective function (1) using the mapped values and the G81 problem instance defined by Ref. [1].

f0d26a05b0316d97de68fb656b3ba8f998d42ef75946e97cd21c138877f2595d343b5a6f9075d8ca6ebc2c8aaadec25fcce1055bb5d09429615b
41c47fa8994971c2b66db7387fbb3a5f59d4d2f83918e039dbe70d0d0b88e3dc04bbd2380aac1f6d1ec0574cb84f549f49bff4514fa958f0c0b2b
0dc22d87ee0dcce9c681765b1164c133dfda8b5a974b6d351a301081289d44875e159ad8652198eaf9b6b094b680b7e2821aa6cbf545795c64
388b6bbea3120b3c4d04fadf01f0ed3bb8162c2f5383cff094ab794212e7b76d5cba1530158472df5de0767d91dd459c1c46433273f841cea579
531b57c172c742331c3120c34770187ee13e4793de741139e678f4572ecb1c9ebe11689c183b83d9b2813eb4732949d82b40659b9a116f7b45
ef893cad5f1e8ef458448732b5e7812ba3bfe4b9178f9abab211489fbf905877444d7bb156aa4d17c8a389420568d55acceb39464acc4919df64
61b9ca45f0ada4690e080aeab6d938092e498efc7ede783b92f30b6a9be89acf385527d05658316235e12dd36223a3bc8f3096326a2e29a1268
8c04509848ee0f08dec557c09463fdcc257187496ce145d8d625e917b7759d59598611fcbe695f894ff185e2ad07e0056783e92c7333646f7bd
86fb935007fe262d0f2494ade17290a471a4554f195e11feb22145f1d7eeb4779da6d740e7aa487966f5fd3a9d248c79b7eb6cf18e30378d157fe
7e8091cf3ab650dd4af7c001b8887a4890cc6b1f80bbc1e0c1b94ebf14aeb902a2b4c3eb540b0d2e1fd0a289baa559f0b9cfacf7e6ecae7f2de05
3337c02dbfdc7cfd07cbde854b75b90b00b68588fbed02e6982f080788247303a8a3a860235d2e1d05ed0edf758d84823619402f773dfc4ca48f
064c32dd76a175263a273f6e36a7a2681c02596271fe634c0a407d63ef27fac48aced0692d98c434aeb35e9b4d95c868226fdd4e49b78917f2a5
bade958e4e204f86ae4a3cf71dd7dd6eecbf4516893628e4471964c5e5dd20e3ff01d48d45dd5d166cea53d8a2297aaf6406b9f0ac03dd3cdd74
540042add7d86d904a317e4fecb8e1f8da83bc5682d2d6270f11c3a78f4d0a5052da9ac76f2f6aa7b99616af0a055e257f641dbd774bed7fc6d3
d3e9a27a3ff8134b047b33ec58ef22d23ffa56cd2169f2d5a99dbfc80dc2d4b02cfdb998f7ac0ce80cd8d12823c81ec77df9ac150bf5b98e04b4a
54a9a05a2a4b376702c81eafce397346524f026369790025ab6abe9869ad0c55efc75074ffde05334dfce11339eca7af8c37fac8b365bb7179330
c93bcc3636ae99f1ca3dbf80aa389dcaa031dae14eda9a4e7d01b501b5cd19ff6a1c3094be1260cd5f8164e658cdc2b5cc4aba55028c43d28e34
119c071c025a9412b37117a54005d80052679be077e68798e9783ee990236b11cde1b7c39b0d121b1fad6743223977955c81ea8ba15166eee1
17a020a0c31db120c1b597fc481ce1fa8b0a36894f693f02b547f0733b8c0a72771933eb913280d02271d881ee15e13db06b3158bb2f08b1584
37f156ea86296a5ed17f615f004d5ad8e1964ad7582d294217c093cc3c2bdf218f4ed6e8baa30a0a7712248e69f211cd99153b551df05b9e2991
f1c5cc343cd0a4722ccb618016981901bc32b5d14818d1db0ae50d39219baca7f9a12bc896c4995b97dda4baa972dedafc0888f8c31ab65237d
bffea534159feff710e77cc5c7ea966ac52d432a47b62fd9f6653303ff75e59ad501d36b11f031dfd70b4abd716be7ebf4c13a3a03a9fc8287c42a
b0bfc8208f0f3940a14a2d6d13692b3bbd368336ac2729e9a0998122cd7a5178daf524a58e5f4741282f555eb82db75f22ea203de1447d22208
d43464c617d71fa3b8a0cf67da7ea19f559b8c9b5974184d383105221c5510a5ba4872b0e16e9225bd463b5091a8bd9b3e27697b2deae37208
eec0bfc19bab71a653ca1d9a720e67037e23db13eec7c6cdb247bc4beb7283f4db756ff8a0a2573a479d5425a2f8214440687fd2eef28f149385
7734f9a2f72eb058c81edce3aadbfbc71b09b74471ffef417e696990cd3c513ca79bbe1c8feabcceacd16082834c59eccada46200ef3edd6a8930
7f8b2859fafb333a52a125dd8c150e26651fb7693f925d800cf869032eb42183514a38aa4c277cf4fec556483a784ae200b315a0f3d2dd7e405e
82abdd5f04d4de8759b40f161ce71f3f9ed03dbb09841856839ec991816ade19f8a75ff0eb7172f1705d7e281a414c1e928f6e07be849e0fe46ec
31cba6c01f128ce395491a0eeff0c8c8dde5c44e505937186735091e63c7348c3514652e6480de8a9036101eee0cb20a28112aafaeb9788956d0
4f1bcfc28aac01081b05e7e91108cffd27208e1abda7c22718a3ef75cf6c6c9aed1922dc3a142c96495dee18bad40025ec5fcdb45200d64016932
bc64cbb286e20461c5a2aec3cb62be3f2c5fad275e4d0bf6fa079e5772a8a6a55e142b972afa924a0a72356e100742e7c81c21dff1c283d734c94
5931cf654ded38cc97b79479020449bc9ba56192715c1cec978f12aa0d0b9de0fbf07d1c5d25de70ab1200c6faa7cfe392d329f39536a6bb57cf
5383c4ebe18f088ea6a879bd548d0c14fa099f7f8eead4f5f112f8c80ff204f1c45706079ad8d57cda1fac767a51496b381d8ddd43b1f15258800
7e3ed9ced0b22869c1070d62fb00c2ab9a12b920ca782b76b11fd31e7c4ea5cad8c8a78118a3624dcaa82107d64326953424c906edaea09b52b
2729f24e85f872f55bd48138b862d730ac4790e2cd1efb10ca0dc004eb3dadc26408f2a954ea2082ad1093a010da2c048360ddf1be9a0229388e
5b414d871a0282b866a2aa8519cfa1df6b25418d96c6fdb8deff69683d6541ef806ca0a3f1899d83eaa2731ad220a1fa8339a8730a702713c0b3
d060e1b968a8a483b8acc1a9df9585229954d9fa1150ef417a09165ca449ae097e23e686761c2fef57858d1cf99b3eb7382cb4bfb8f598e0eb79f
5026e8ab43221037051eb516338577c64520f82fcd46e8ebfccc614b7577f670f273424ad4f71f4dcdf1c94df4502a05f60f9e605477e40b4bf21
b3e41d7bf9319927b1a671dc27bda4109c64bc2e7e918163ef5566350a63ebf4b59454f2331e4360a9aa91a7186a3a637d09369cd250d091a13
76100260001821d2d74b3b2561ea41249b5344f5091a9b443bdb24828707358dfb74919678e2a803871b9b393b459eb35aaf366de397964e0
c21c8a72450f0aa8b30cf82b7ffd5cc0e40314eef4b7cefa1f462d52a11d55a2d21d286b9bb3176971761ba31d8cffda173b890ed26cac14343cc
78b3d2c243b0b5b5b8a9288bf0e777e7a821f012b16bc30b165823cc9cc9c9bc87e975bc676559ea99df33ebeae741198649555d4dd7767989
19e969cc0720b2092295f8ec1bb6239d2d78e8314f741a432e0fae10372b311e6703da56a873f8575b476101da8cfcb53bba9ccf6e6c9200b93c
d3528ae8f7f470fbc